\documentclass[]{raa}            
\usepackage{graphicx,times}
\usepackage{natbib}
\usepackage{amssymb}

\providecommand{\bm}[1]{\mbox{\boldmath$#1$\unboldmath}}
\def\kms{$\rm{km~s}^{-1}$}

\begin{document}

   \title{ Star formation properties of galaxy cluster A1767 }

 \volnopage{ {\bf 2015} Vol.\ {\bf X} No. {\bf XX}, 000--000}
   \setcounter{page}{1}

   \author{Peng-Fei Yan
      \inst{1,2}
    \and Feng Li
      \inst{1,3}
   \and Qi-Rong Yuan
      \inst{1}
   }

   \institute{Department of Physics, Nanjing Normal University,
                 WenYuan Road 1, Nanjing 210023, China;
                 {\it  pfyan0822@sina.com; yuanqirong@njnu.edu.cn}\\
            \and
                 School of Mathematics and Physics, Qingdao University of Science and Technology,
                 SongLing Road 99, Qingdao 266061, China
            \and
                 School of Mathematics and Physics, Changzhou University, GeHu Road 1, Changzhou 213164, China
\vs \no
}

\abstract{Abell 1767 is a dynamically relaxed, cD cluster of
galaxies with a redshift of 0.0703. Among 250 spectroscopically
confirmed member galaxies within a projected radius of
$2.5r_{200}$, 243 galaxies ($\sim 97\%$) are spectroscopically
covered by the Sloan Digital Sky Survey (SDSS). Based on this
homogeneous spectral sample, the stellar evolutionary synthesis
code, STARLIGHT, is applied to investigate the stellar populations
and star formation histories (SFHs) of cluster galaxies. The star
formation properties of galaxies, such as mean stellar ages,
metallicities, stellar masses, and star formation rates (SFRs),
are presented as the functions of local galaxy density. Strong
environmental effect is found in the manner that massive galaxies
in the high-density core region of cluster tend to have higher
metallicities, longer mean stellar ages, and lower specific star
formation rates (SSFRs), and their recent star formation
activities have been remarkably suppressed. In addition, the
correlations of the metallicity and SSFR with stellar mass are
confirmed.  }

   \authorrunning{P.-F. Yan et al. }    
   \titlerunning{ Star formation properties of galaxy cluster A1767 }  
   \maketitle
\section{Introduction}           
\label{sect:intro}
Clusters of galaxies, which are the largest
gravitationally bound systems in the universe, have long been
recognized as entities that can provide vital clues to large-scale
structure formation. Nowadays, the evolution of galaxies in dense
environment has become an important research field in
extragalactic astrophysics. The star formation properties of
cluster galaxies and their correlations with galaxy morphology,
environment and some other physical quantities contribute to our
understanding of galaxy formation and evolution
(\citealt{kennicutt98,brinchmann04,kauffmann04}). The star
formation histories (SFHs) of member galaxies may shed some light
on the evolution of their host cluster. As a dynamical tracer of
luminous matter in galaxy clusters, member galaxies provide
wealthy information in their high-resolution spectra for analyzing
dynamical structures of clusters, the SFHs and chemical content of
the galaxies.

In recent years, with its enormous amount of homogeneous spectroscopic and photometric data,
the Sloan Digital Sky Survey (SDSS, \citealt{york00, stoughton02};
\citealt{abazajian03, abazajian04})
has provided more samples of galaxies for studying the star
formation activities of the galaxies in different gravitational
environments (\citealt{kauffmann04, brinchmann04,
asari07,deng13,ricciardelli14}).

Star formation rate (SFR) is an important physical parameter
indicating the intensity of star formation activities in galaxies.
The phenomena of HII 
regions, OB associations and starbursts are all significant
indicators of current star formation within galaxies. The galaxy
SFH refers to how the SFR evolves with cosmic time. And the
mixing-ratio of various stellar populations can, to a certain
extent, reflect the SFHs of galaxies. Certainly, there exist many
factors that are closely related to galaxy SFR, for example, the
physical properties of individual galaxies (e.g., luminosity,
mass, gas content, morphological type, etc)
(\citealt{kennicutt98,kauffmann04}), and environmental influence
(e.g., tidal forces, gas stripping, strangulation, strong galaxy
interactions and mergers, etc)
(\citealt{dressler80,rasmussen12,ideue13}). On the whole, at
present late-type galaxies are usually forming stars more actively
than early-type galaxies, SFR in cluster galaxies is lower than in
their field counterparts, and low-mass galaxies are undergoing
more violent star formation activities than those massive galaxies
(\citealt{kennicutt98,kauffmann04,hern13,wagner14}). Following the
standard hierarchical clustering scenario of cosmological
large-scale structure, galaxy clusters formed through successively
accreting of surrounding field galaxies. Once field galaxies
access dense regions (e.g. galaxy clusters), their star formation
activities would be gradually suppressed due to manifold physical
mechanisms (\citealt{balogh02,koopmann04}). Additionally, there
are clues that star formation activity is also suppressed in some
high-redshift clusters of galaxies (\citealt{cid05}).

The nearby ($z=0.0703\pm 0.0004$, \citealt{oegerle01}) Abell
cluster, A1767, located at
13$^\mathrm{h}$36$^\mathrm{m}$00$^\mathrm{s}$.3,
+59$^{\circ}$12$'$43$''$ (J2000.0), is a  cD galaxy cluster with
the BM-type II and richness $R=1$ (\citealt{abell89}).
The predominant galaxy is the central cD galaxy, MCG+10-19-096 (\citealt{hill98}).
Using the spectroscopic data of 58 member galaxies,
\citet{hill98} derived the mean heliocentric velocity of
$21,069\pm121$ \kms, and the velocity dispersion of
$849^{+92}_{-70}$ \kms. They also pointed out noticeable lacks of
emission-line galaxies in A1767, compared with other clusters.
Furthermore, the {\it I}-statistics showed that the velocity
distribution of A1767 is Gaussian (\citealt{oegerle01}), and there
is no statistically significant substructure as indicated by the
$\mathrm \Delta$-test in \citet{dressler88}, showing that A1767 is
a nearly dynamically relaxed galaxy cluster
(\citealt{hill98,wojtak10}). Analysis of spatial orientations of
the galaxies in A1767 showed that the spin vectors of galaxies
tend to lie in local supercluster plane and the projections of the
spin vector tend to be oriented perpendicular with respect to the
Virgo cluster centre (\citealt{aryal07}). Based on 147
spectroscopic data, \citet{wojtak10} derived the virial mass of
galaxy cluster A1767 is
$11.84^{+1.40}_{-3.60}\times10^{14}\mathrm{M}_{\odot}$, which is
basically consistent with the weak lensing survey
(\citealt{kubo09}) and the result of \citet{popesso07}. They also
judged that there is no cool core in A1767 on the basis on
\citet{hudson10}. Using the $g$ and $r$ photometry from the SDSS,
\citet{andreon10} found that stellar mass of A1767 within virial
radii ($r_\mathrm{200}$) is $10^{12.87\pm0.06}$
$\mathrm{M}_{\odot}$. \citet{plionis09} analyzed 159 X-ray data of
A1767 and found the X-ray temperature and luminosity are 4.1 keV
and $2.43\times10^{44}$ erg~s$^{-1}$, respectively, in agreement
with the X-ray temperature revealed by \citet{lin04} and the X-ray
luminosity derived by \citet{rudnick09}. Using the SDSS data
within a projected radius of $r_{200}$, \citet{crawford09} studied
the luminosity function of A1767 in $B$ band, and found the
characteristic magnitude is $M^*_{B}=-20.24^{+0.36}_{-0.4}$ .
\citet{poggianti06} analyzed 48 SDSS spectra within a projected
radius $r_\mathrm{200}$ of A1767, and derived that the fraction of
the star-forming galaxies with a significant emission line OII at
3727 \AA , a reliable signal of ongoing star formation, is about
$0.19\pm0.06$.

This paper aims to unveil the SFHs of the galaxies with various
morphologies in A1767. The rest of this paper is organized as
follows. The SDSS spectroscopic data of A1767 and the description
of spectral fitting method, STARLIGHT, are presented in Section 2.
In Section 3, we analyze the SDSS spectra for 243 member galaxies,
and derive their star formation properties, such as SFRs, mean
stellar ages, stellar metallicities, and their variations with
local galaxy density as well as the stellar mass assembled in
cluster galaxies. Finally, some conclusions are given in Section
4. Throughout this paper, We assume a flat cosmology with $H_0=70$
\kms Mpc$^{-1}$, $\Omega_m=0.3$, and $\Omega_\Lambda=0.7$.

\section{ Data and analysis }
\label{sect: Data}

For investigating the star formation properties of cluster
galaxies as a function of local density, a larger radius,
2.5$r_{200}$, will be adopted in this paper, in order to cover a
wider range of local galaxy density.  $r_{200}$ is defined as
where the mean interior density is 200 times the critical density,
which can be derived from the cluster redshift ($z_{c}$) and
velocity dispersion ($\sigma_c$) (\citealt{carlberg97}):

$$r_{200}=\frac{\sqrt{3}\sigma}{10H(z_c)},\,\,
{\rm
where}\,\,H^2(z_c)=H^2_0[\Omega_m(1+z_c)^3+\Omega_{\Lambda}].$$

We initially take $z_c=0.0703$ and $\sigma=849$ \kms , given by
\citet{hill98}, and achieve $r_{200}= 2.13$ Mpc. Considering a
scale of 1.342 kpc/arcsec for A1767, 579 normal galaxies with
known spectroscopic redshifts ($z_{\rm sp}$) within a projected
radius of 66.1 arcmin (i.e., $2.5r_{200}$) are extracted from the
NASA/IPAC Extragalactic Database (NED). The rest-frame velocity
for each galaxy can be computed by $V=c(z-z_c)/(1+z_c)$. We take a
standard iterative 3$\sigma$-clipping algorithm
(\citealt{yahil77}) for selecting member galaxies, and apply the
ROSTAT software (\citealt{beers90}) to calculate two resistant and
robust estimators, namely the biweight location ($C_{\rm BI}$) and
scale ($S_{\rm BI}$), which are analogous to the mean value and
the standard deviation. The biweight location ($C_{\rm BI}$) is
taken for correcting the mean redshift of A1767, and scale
($S_{\rm BI}$) is adopted as velocity dispersion ($\sigma$). Then,
we re-calculate the scale and virial radius $r_{200}$, and find a
smaller $r_{200}$. Those galaxies with projected cluster-centric
distance $R>2.5r_{200}$ are excluded, and the iterative
3$\sigma$-clipping algorithm can be applied again for the updated
sample of remaining member galaxies. This is an iterative process
for achieving the final values of mean redshift ($z_c$), velocity
dispersion ($\sigma$), and virial radius ($r_{200}$) for A1767. As
a result, we obtain a sample of 250 member galaxies with mean
redshift of $z_c=0.0711$. The velocity dispersion is
$\sigma=803\pm62$ \kms, and its corresponding dynamic radius is
$r_{200}=1.922$ Mpc. These galaxies are distributed within a
circular region of $R=2.5r_{200}=4.805$ Mpc.


Among the 250 galaxies with $0.0621 < z_{\rm sp} < 0.0801$, 243
($\sim 97\%$) member galaxies are covered by the SDSS
spectroscopy. The left panel of Fig.~1 shows the spatial
distribution of member galaxies in A1767, and the 243 galaxies
having SDSS spectra are denoted by circles. Although the selection
effects of SDSS spectroscopic survey, to a certain extent, affect
the completeness of the galaxy sample, this homogeneous spectral
sample seems to have a completeness of $\sim$ 90\%
(\citealt{blanton05}). As we know, the SDSS main spectroscopic
galaxy sample is complete within the magnitude range $14.5 < r <
17.77$. For the member galaxies in A1767, their absolute magnitude
are in the range $-23.03 < M_r < -19.76$, only 8 of 243
spectroscopic galaxies are beyond this scope.  As for the fibre
collision problem, which is due to the fact that fibres cannot be
placed closer than 55$''$, is responsible for the most
incompleteness in the SDSS data. \citet{strauss02} estimated that
affects $\sim$ 6 per cent of all target galaxies. Spectroscopic
fibers have been assigned to objects on the sky using an efficient
tiling algorithm designed to optimize completeness
(\citealt{blanton03}).  These SDSS member galaxies are scattered
in whole cluster region, and they can be used to trace various
density environments within A1767, which is suitable for
statistical research on the variation of star formation property
with local galaxy density. The right panel of Fig.~1 present
distribution of the rest-frame velocities of cluster galaxies.

\begin{figure}[!h]
\centering
\includegraphics[width=62mm]{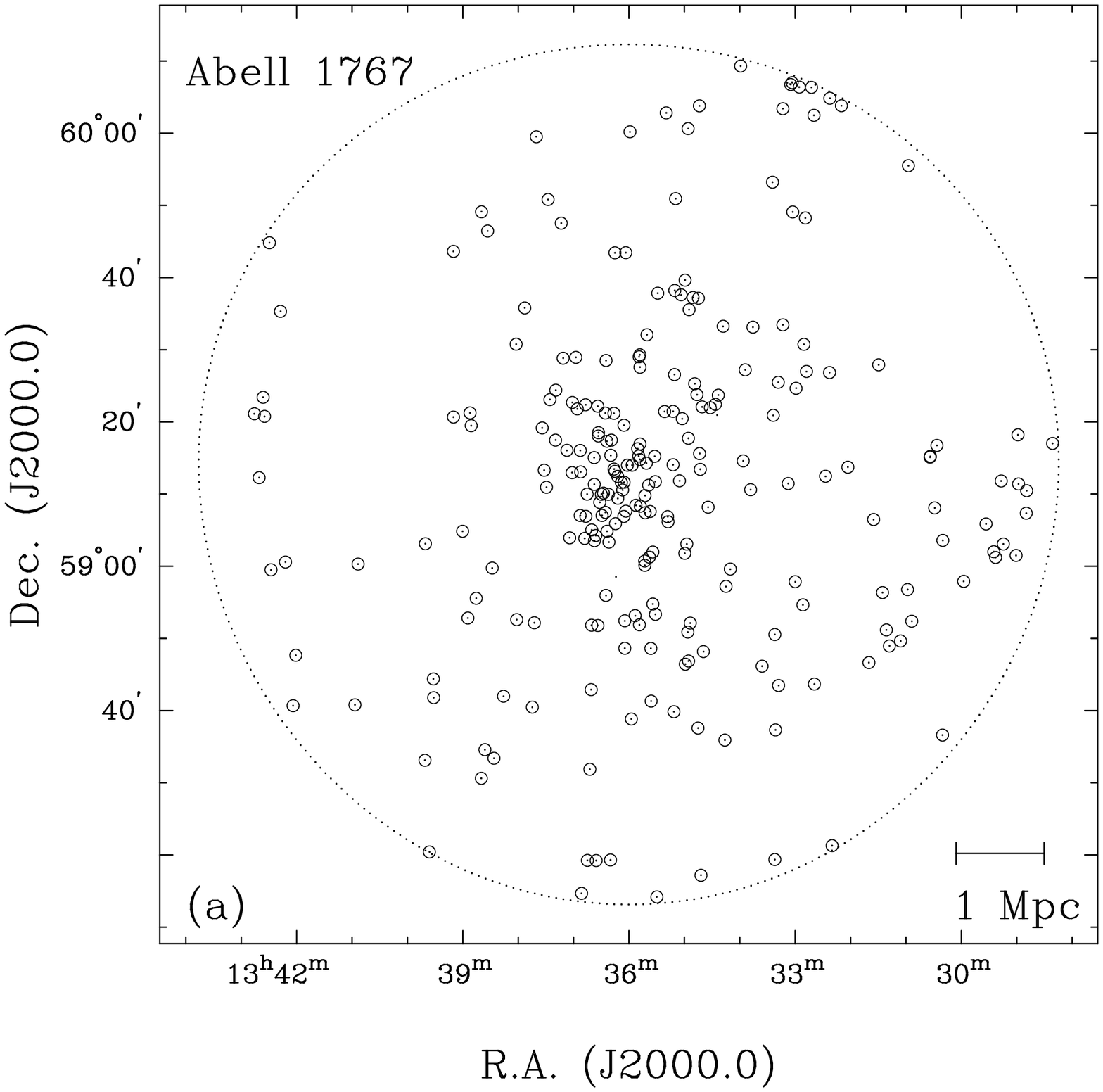}
\includegraphics[width=62mm]{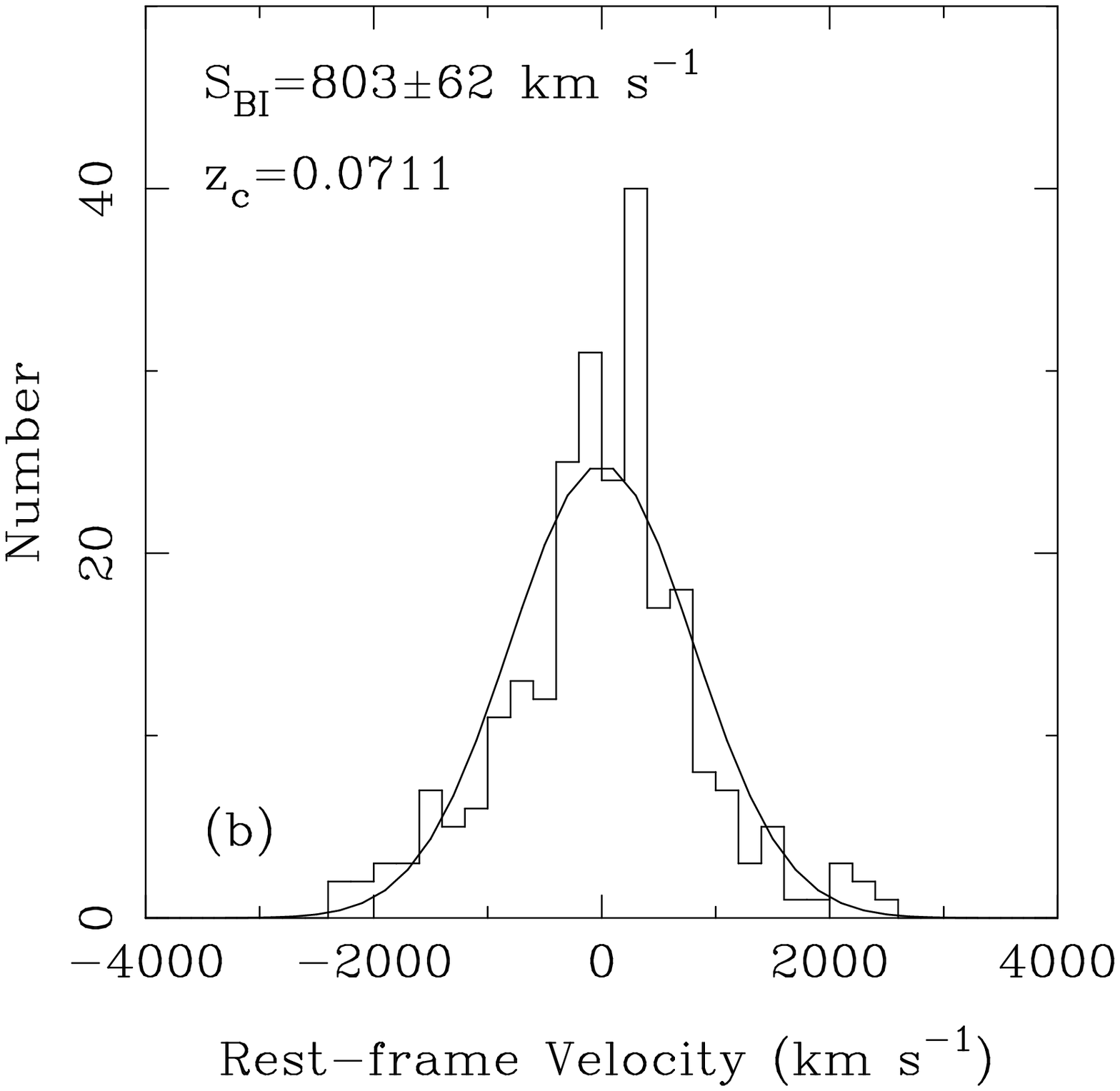}
\caption{(a)Spatial distribution of member galaxies of Abell 1767.
The points represent 250 member galaxies,  while the circles
denote the 243 SDSS member galaxies. The A1767 region defined by a
radius of 2.5$r_{200}$ is also shown. (b) Distribution of the
rest-frame velocities of member galaxies}
 \label{fig1}
\end{figure}


Simple stellar populations (SSPs) refer to the simultaneous
generations of stars with same stellar metallicity for a specified
initial mass function (IMF). Based on the SDSS photometric data
and the star formation parameters derived by MPA/JHU group
(\citealt{brinchmann04,kauffmann04}), \citet{yuan05} analyzed the
star formation properties of 184 member galaxies of different
morphological types in A2199.
In this paper, we use the spectral synthesis code, STARLIGHT
(\citealt{cid05}), to decompose the observed galaxy spectra of
A1767 in terms of a linear superposition of a series of SSPs of
various ages and metallicities. The star formation and chemical
histroies for cluster galaxies can be produced as output. It
should be noted that the STARLIGHT only fits the regions of galaxy
continuums and absorption lines, bad pixels and the windows of
emission lines and the Na D doublet are masked and left out of the
fits. The emission-line masks are constructed in a
galaxy-by-galaxy basis. Following \citet{asari07}, we can
generally deduce the SFHs of galaxies on the basis of the output
SSP series at different epochs. We make use of a base of 150
($N_{\star}$) SSPs extracted from the stellar population
evolutionary synthesis models of \citet{bc03} for a Chabrier IMF
(\citealt{chabrier03}), spanning 25 ages (1 Myr$<t_{\star}<$
18Gyr) and 6 metallicities ($0.005 \leqslant Z_{\star} \leqslant
2.5 Z_{\odot}$). Each SSP with age $t_{j}$ and metallicity $Z_{j}$
contributes a fraction of $x_{j}$ to model flux at a chosen
normalization wavelength ($\lambda_{0} = 4020 {\rm \AA}$), and its
contribution can be equivalently expressed as a mass fraction
vector $\mu_{j}$. As STARLIGHT outputs, we obain mean stellar
ages, mean metallicities, present-day stellar mass, as well as the
full time-dependent star formation and chemical evolution
histories. Therefore the current ($\tau < 24.5$ Myr) SFRs and
specific SFRs can be derived for cluster galaxies. Besides,
STARLIGHT outputs intrinsic extinction $A_V$, velocity dispersion
$\sigma$ and some other parameters. The star formation parameters
used for further statistical analysis in our work are all derived
from the STARLIGHT outputs.

\section{ Statistical results }

This work focuses on the variation of star formation properties
(e.g., SFR, mean stellar age, stellar metallicity and stellar
mass, etc) of member galaxies with local galaxy density. To
characterize the density environment, we use the sample of 250
member galaxies to estimate local surface density ($\Sigma$).
Following \citet{dressler80}, we define $\Sigma$ by the nearest 10
neighboring galaxies centered on each member galaxy,
$\Sigma_{10}=10/(\pi d^2_{10})$, where $d_{10}$ is the projected
distance to the 10th nearest neighbor. Thus, $\Sigma_{10}$ is not
constrained by the cluster size, shape and other overall
properties.  Considering the completeness of SDSS
spectroscopy, the derived $\Sigma$ values should be about 10 \%
smaller than intrinsic local galaxy densities.

The mean stellar age is an essential parameter characterizing the
stellar population mixture of a galaxy. Compared with the
light-weighted mean age, the mass-weighted mean age should be more
intrinsic in terms of what that physically means. The same applies
to the mean metallicity. As described in \citet{cid05}, the
mass-weighted mean age and mean stellar metallicity can be derived
by
$$ \langle \log t_{\star}\rangle _{M} =
\sum \limits_{j=1}^{N_{\star}}\mu_{j} \log t_{\star, j}, \,\,\,\,
 {\rm and} \,\,
 \langle Z_{\star} \rangle _{M} = \sum
\limits_{j=1}^{N_{\star}} \mu_{j}Z_{\star,j},
$$
where $t_{\star, j}$ and $Z_{\star, j}$ denote age and metallicity
of each SSP component, respectively.

Several criteria have been proposed to distinguish the
early-type (red) galaxies based on the SDSS data, such as
brightness fraction of de Vaucouleurs component ($\rm{fracDeV} >
0.5$), color index ($u-r
> 2.2$), and spectral features (such as continuum slope,
Mg, Ca II H and K absorption lines, H$\alpha$ emission, etc).  It
should be noted that the criterium $\rm{fracDeV}>0.5$ probably
fails for the spiral galaxies with prominent bulges, while the
color cut $u-r>2.2$ probably misclassifies the star-forming
galaxies with strong H$\alpha$ emission. For achieving a reliable
morphology classification, we inspected all images and spectra of
these 243 member galaxies one by one, and balanced the values of
\texttt{fracDeV} and $u-r$. As a result, 159 member galaxies are
classified as early-type galaxies, and 84 are classified as the
late-types. For verifying the reliability of morphology, the
color-magnitude relation for A1767 is shown in Fig.~2. A linear
fit for the early-type galaxies is shown by a solid line. The
dashed-dotted line corresponds to the SDSS spectroscopy
completeness.

\begin{figure}[!b]
\centering
\includegraphics[width=70mm,height=70mm]{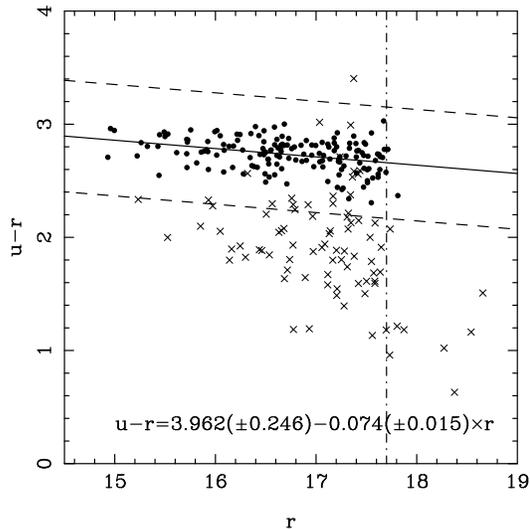}
\caption{ \baselineskip 3.6mm The color-magnitude relation for the
member galaxies in A1767. The early-type galaxies are denoted by
the spots, and the remaining late-types are denoted by the
crosses. A linear fit for the early-type galaxies is shown by a
solid line, and dashed lines correspond to $2\sigma$ deviation.
The dashed-dotted line corresponds to the SDSS spectroscopy
completeness.} \label{fig2}
\end{figure}

Fig.~3 presents the mass-weighted mean stellar ages and mean
metallicities as functions of local galaxy density. The slope of
our linear fitting and the Spearman correlation coefficient are
also given in the plots. Fig.~3 reveals a significant increasing
trend of mean ages with galaxy density, and the mean metallicities
also appear a similar slight trend. It shows that the dense core
region of A1767 is predominated by early-type galaxies, consistent
with the morphology-density relation first pointed by
\citet{dressler80}. Basically, early-type galaxies have older
stellar ages than late-types. The SFHs of early-type galaxies show
that they formed earlier, and there are no signs of star formation
in the recent 1 Gyr, which leads to rather old age of the overall
stellar population.


\begin{figure}[!t]
\centering
\includegraphics[width=67mm]{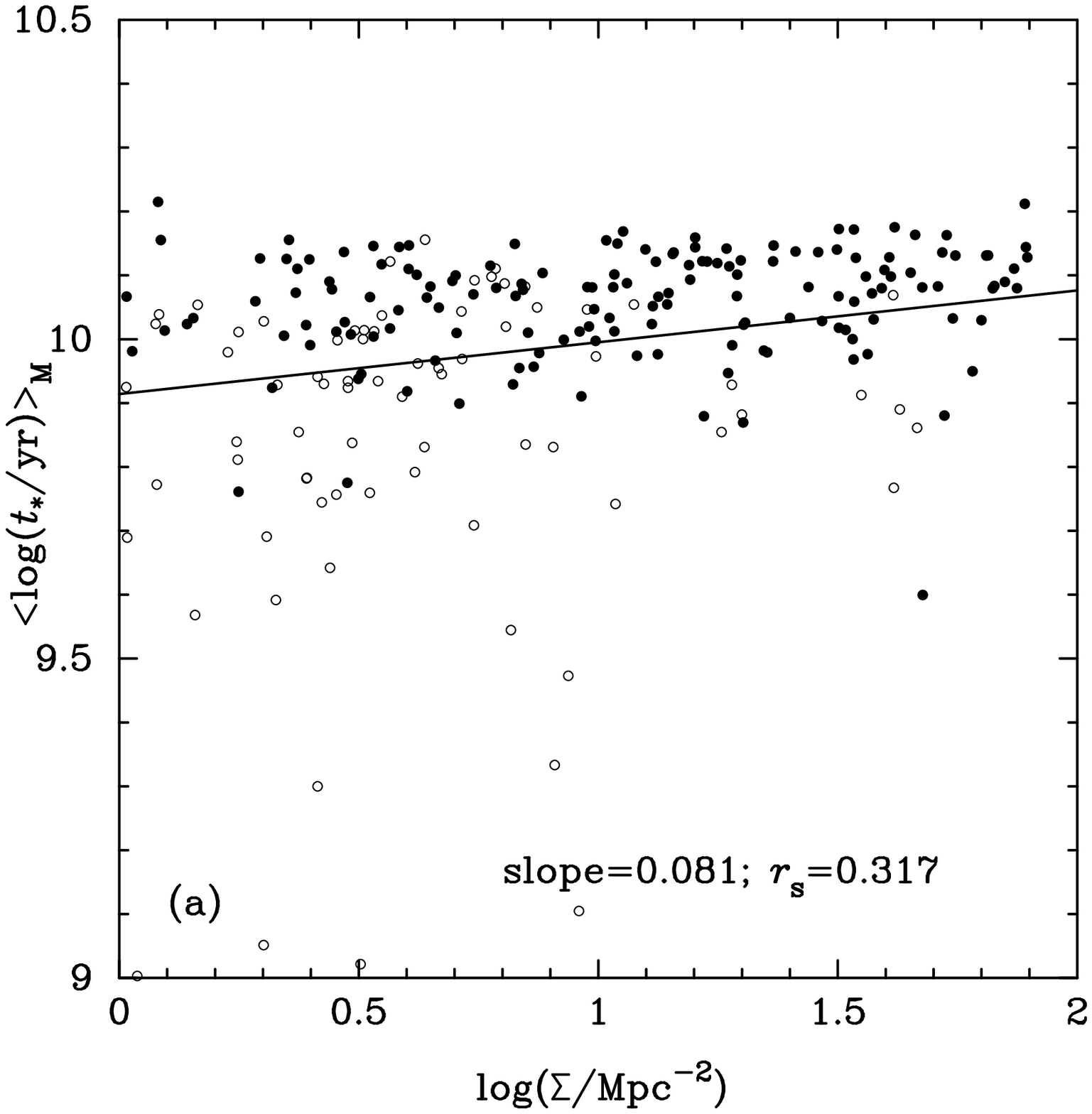}
\includegraphics[width=67mm]{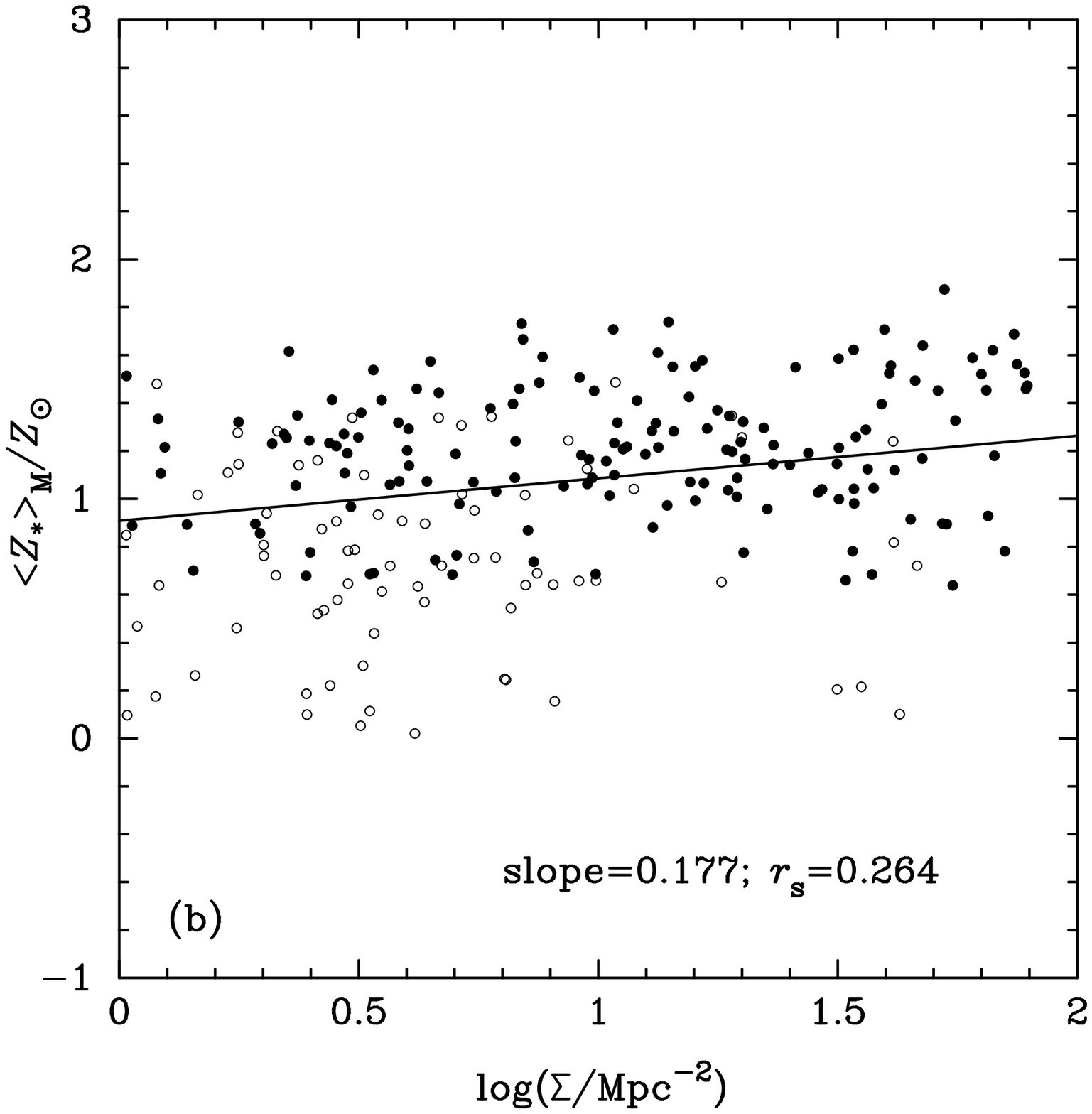}
\caption{ \baselineskip 3.6mm  Distributions of the mass-weighted
mean stellar ages (a) and mean metallicities (b) for 243 SDSS
galaxies along local galaxy densities. Solid lines indicate the
linear fittings.
Early-, late-type member galaxies are denoted by solid and open
circles, respectively.} \label{fig3}
\end{figure}

\begin{figure}[!b]
\centering
\includegraphics[width=70mm,height=70mm]{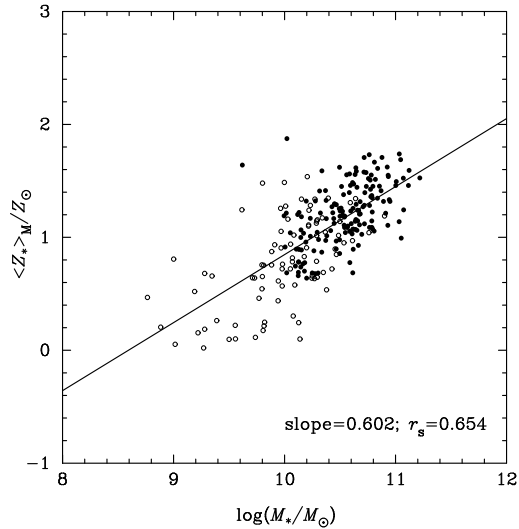}
\caption{ \baselineskip 3.6mm Correlation between the
mass-weighted mean metallicity and stellar mass for 243 member
galaxies. A linear fitting is also given.} \label{fig4}
\end{figure}

The right panel of Fig.~3 presents that the mean galaxy
metallicities are weakly correlated with local density, which is
mainly due to the dependence of metallicity upon stellar mass.
Fig.~4 exhibits the correlation between the mean stellar
metallicities and the stellar mass assembled in galaxies. It can
be seen that the metallicities have a large scatter for low-mass
galaxies, while the massive galaxies with $M_{\star}>10^{10}
M_{\odot}$ are mainly early-type galaxies, and their metallicities
correlate strongly with stellar masses. It means that more massive
galaxies have higher metallicities,
which can be interpreted in terms of a known mass-metallicity
relation (e.g. \citealt{garnett87,tremonti04}).
It is noteworthy that most studies (e.g. \citealt{tremonti04})
measured the metallicity by means of the emission line luminosity,
which represent the metallicity of interstellar medium (ISM).
Nevertheless, the metallicity in this work is derived by
accumulating the best-fitting SSP sequence, thus it represents the
stellar metallicity. Considering that the ISM metals mainly
originated from the feedback during the evolution of stars (such
as galactic wind, supernova explosion, etc), and the stars
successively formed certainly retain the metallicity of local ISM.
It can be expected that chemical enrichment levels in stars and
ISM scale with each other, and they should follow similar stellar
mass-metallicity relation. If the mass fraction of stars in a
galaxy remains stable, then this relation implies that the
low-mass galaxies with shallow potential wells are difficult to
prevent the heavy metals from being removed by the galactic winds
(\citealt{tremonti04}), which leads to the decrease in
metallicity. A certain dispersion is presented in the stellar
mass-metallicity correlation, and the physical properties of
galaxies (e.g. the galaxy morphology, color, inclination, the mass
proportions of stars and gas within a galaxy, etc) and local
density environment may be the causes of dispersion in such
correlation. The metallicity dispersion is especially larger for
low-mass galaxies, which may be due to the lower spectral
signal-to-noise ratios (S/N) that leads to the more serious
degeneracies among stellar age, extinction and metallicity.

\begin{figure}[!b]
\centering
\includegraphics[width=62mm,height=62mm]{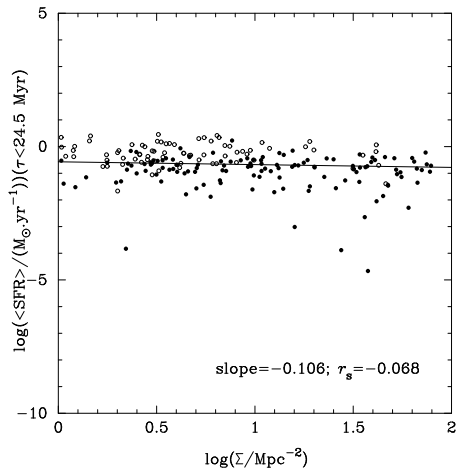}
\caption{ Distribution of the recent ($\tau < 24.5$ Myr) mean SFR
along local surface density.} \label{fig5}
\end{figure}

Current SFR is commonly measured  by the H$\alpha$ luminosity, and
also can be drived from the SFH produced by our stellar population
synthesis analysis. The recent mean SFRs of galaxies, $\langle
{\rm SFR} \rangle$ ($\tau < 24.5$ Myr), yields the best
correlation with the H$\alpha$-derived SFR, where $\tau$ is the
look-back time (\citealt{asari07}). Fig.~5 shows the relation
between recent mean SFR and local surface density. We find that
the star formation activities in massive galaxies have been
seriously restrained in dense regions, and their recent SFRs are
relatively low; while the low-mass galaxies in outer sparse
regions still keep relatively active star formation.

To objectively characterize the intensity of star formation
activities for galaxies with different masses, SFR is usually
normalized on the total mass converted to stars over the galaxy
history until $t_{\star} = 0$, and the quantity ($\rm
SFR/M_{\star}$) is called {\it specific} SFR (SSFR)
(\citealt{asari07}), which measures the pace at which star
formation proceeds with respect to the mass already converted into
stars. Fig.~6 presents the SSFRs of member galaxies as a function
of local surface density. As can be seen from the figure, the mean
galaxy SSFR over the whole history of star formation decreases
with increasing local surface density, indicating that the star
formation activity should be constrained in high-density
environments.

\begin{figure}[!t]
\centering
\includegraphics[width=62mm,height=62mm]{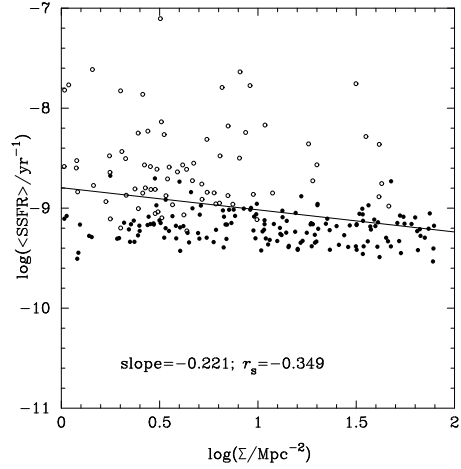}
\caption{ Distribution of the mean SSFRs during the whole histroy
of star formation along local surface density.} \label{fig6}
\end{figure}

\begin{figure}[!b]
\centering
\includegraphics[width=62mm]{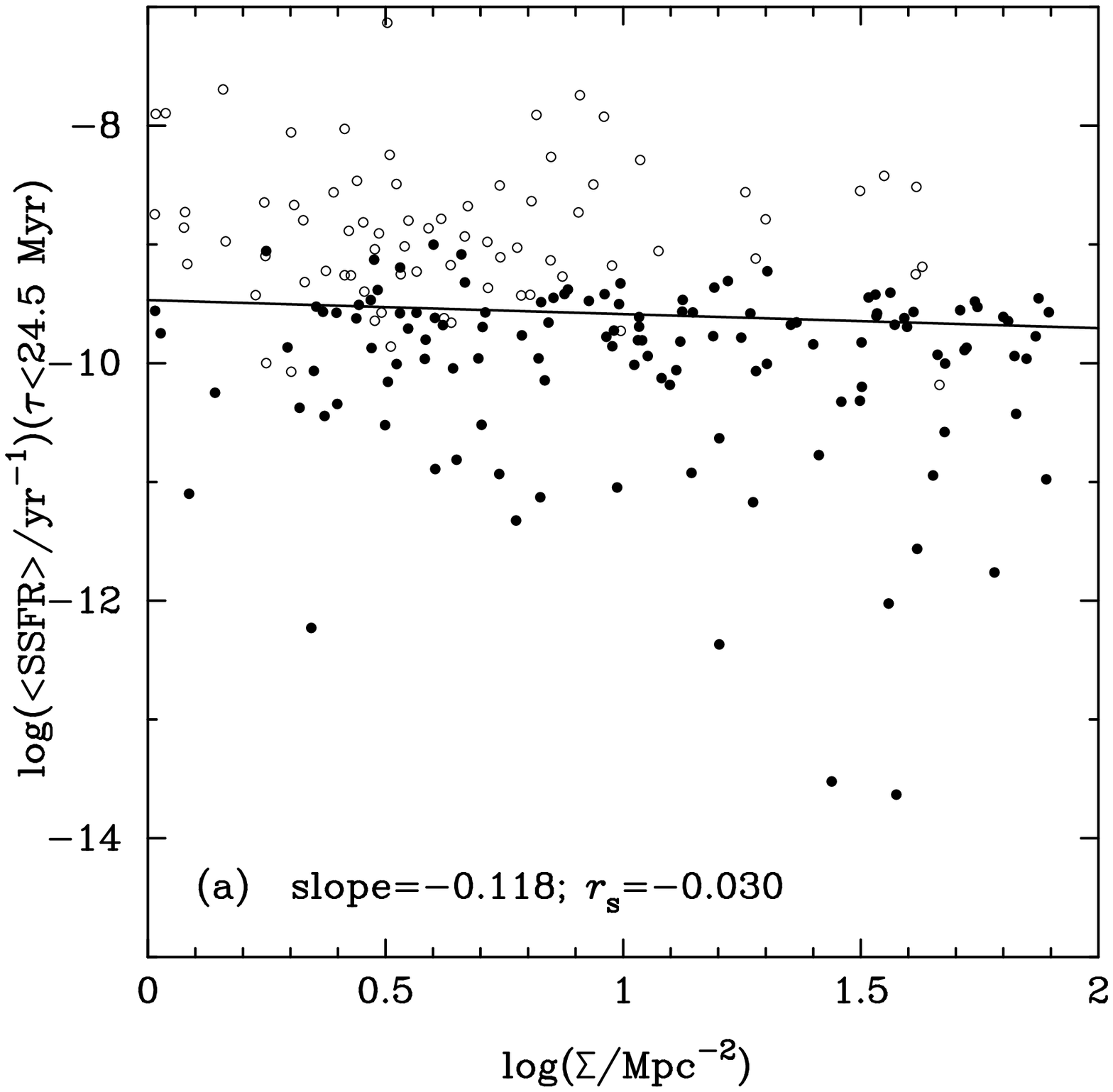}
\includegraphics[width=62mm]{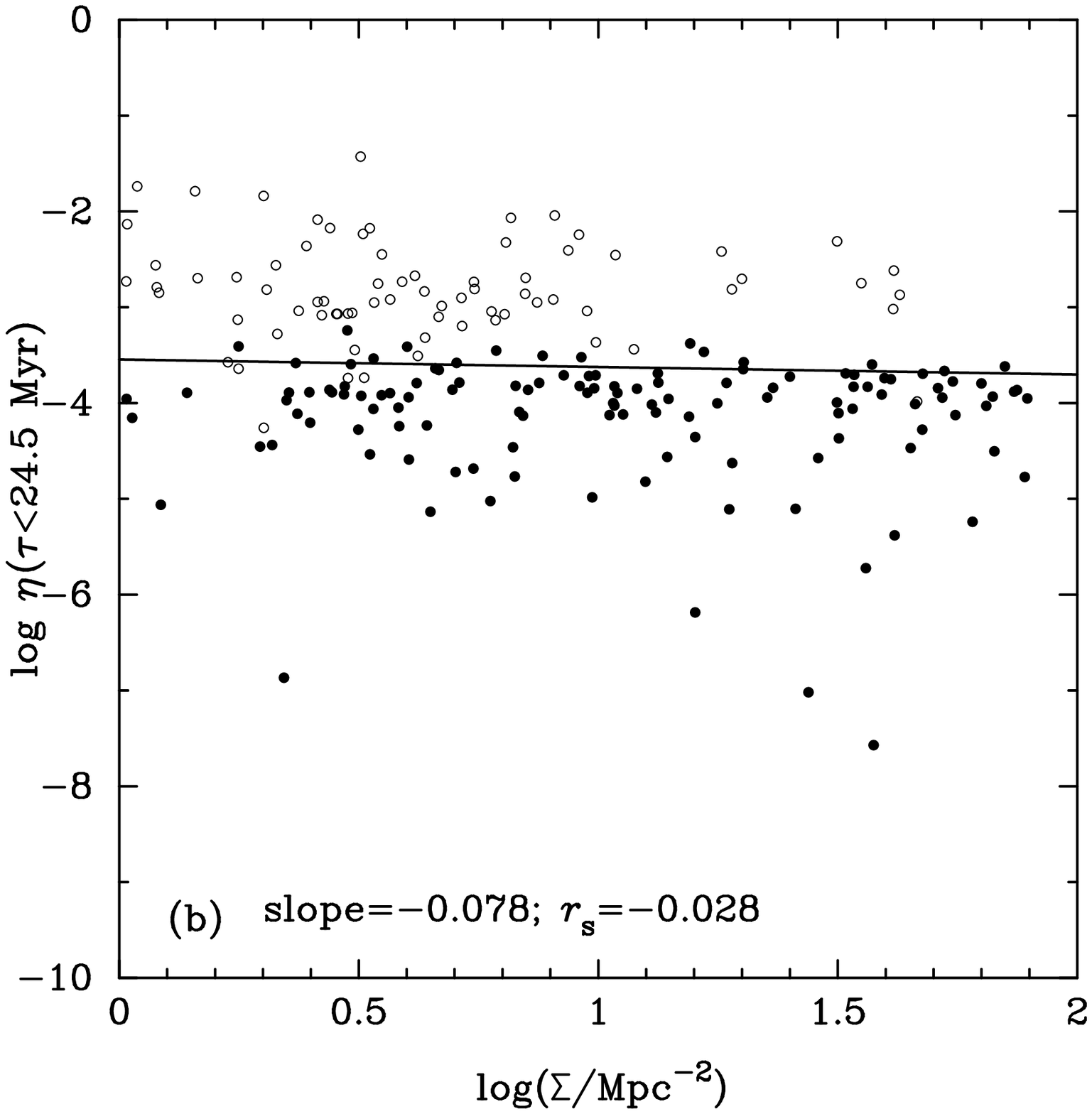}
\caption{ \baselineskip 3.6mm Distribution of (a) the mean SSFRs
and (b) stellar mass fraction ($\eta$) within recent 24.5 Myr
along local surface density. } \label{fig7}
\end{figure}

In addition, we can also derive what fraction of the total
present-day stellar mass ($M_{\star}$) was converted to stars
during recent period ($\tau < 24.5$ Myr) by formula $\eta =
\sum_{t_{\star,j}<\tau} \mu_{j}$ (\citealt{asari07}).
As shown in Fig.~7, both the recent mean SSFRs and mass fraction
of recently formed stars ($\eta$) are indicators of the intensity
of recent star formation in galaxies, and they show similar trends
with local surface density: the galaxies in low-density outer
regions still maintain high degree of star formation over recent
period. On the other hand, for the galaxies in cluster core
region, their recent star formation activities are no more active,
evidently affected by the environmental effects. According to the
hierarchical scenario of cosmic large-scale structure, when the
outer galaxies fall into the core region, the interaction between
galaxy and intra-cluster medium (ICM) will be significantly
enhanced, and the physical processes such as the tides,
galaxy-galaxy interaction, harassment, strangulation, cannibalism,
and dynamical friction (\citealt{poggianti04,yuan05}) would stripe
or take away the gas in galaxies, thus their star formation
activities would be gradually reduced, until slowly quenched.

\begin{figure}[!t]
\centering
\includegraphics[width=62mm,height=62mm]{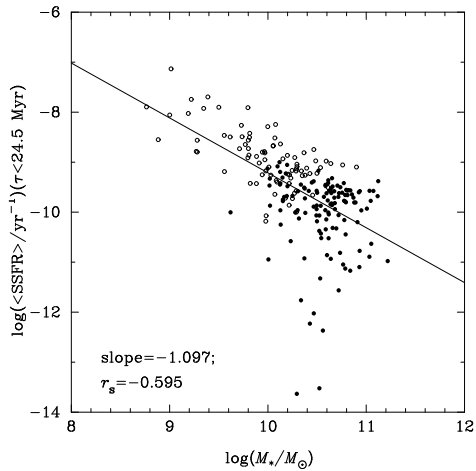}
\caption{ Relation of the galaxy mean SSFRs within recent 24.5 Myr
with their assembled stellar masses.} \label{fig8}
\end{figure}

Fig.~8 presents the variation of galaxy $\langle \rm SSFR \rangle$
($\tau < 24.5$ Myr) with their stellar masses assembled. A strong
correlation is clearly seen, although with some dispersion.
Massive galaxies have smaller mean SSFRs in recent 24.5 Myr, which
can be well explained in the context of hierarchical cosmological
scenario (\citealt{poggianti04}), implying that with increasing
stellar mass, gases are gradually consumed and their mass
proportions of recently-formed stars become even lower, which
accounts for the decrease in recent SSFRs.

\section{Conclusion}

A1767 is a dynamically relaxed nearby galaxy cluster. Within a
projected radius of $R=2.5r_{200}=4.805$ Mpc, a sample of 250
spectroscopically confirmed member galaxies is selected. Our
analysis of this sample shows that the mean cluster redshift is
0.0711, and velocity dispersion is $803 \pm 62$ \kms. 243 galaxies
are found to have SDSS spectra, which provide a homogeneous
spectral sample for studying the stellar population and star
formation histories for the cluster galaxies. The stellar
population synthesis code, STARLIGHT, is applied to obtain the
star formation properties for each galaxies. This work focuses on
how the star formation properties of galaxies (e.g. mean stellar
ages, metallicities, stellar masses, and recent specific SFR, etc)
vary with local galaxy density.
We find that the galaxies in dense core region of cluster tend to
have older mean stellar ages, higher metallicities and lower SFRs,
and their star formation activities are significantly restrained
in recent 24.5 Myr. Besides, we also confirm the correlations of
galaxy metallicities and SSFRs with stellar masses. Environmental
effects on the star formation properties of the galaxies
in A1767 are basically consistent with our previous studies on the
spectral energy distributions for cluster galaxies, where we used
different observation data and analysis methods for the nearby
clusters at different dynamical status. Our cluster sample
includes the dynamically complex clusters, namely A2255
(\citealt{yuan05}), A98 (\citealt{zhang10}), A119
(\citealt{tian12}), A671(\citealt{pan12}), and A2319
(\citealt{yan14}), and dynamically relaxed cluster, namely A2589
(\citealt{liu11}).
Recently, we carried out a similar
investigation based on the same spectral source and synthesis code
for the merging cluster A85 (\citealt{yuan14}). 
Comparison
between these works indicates that the statistical regularities
seem to be weakly related to dynamical stages of cluster
evolution, and might correlate strongly with stellar mass and
local galaxy density.

\normalem
\begin{acknowledgements}

This work is funded by the National Natural Science Foundation of
China (NSFC) (Nos. 11173016, 11433005) and by the
Special Research Found for the Doctoral Program of Higher
Education (grant No. 20133207110006). We would like to thank Prof.
Xianzhong Zheng at Purple Mountain Observatory for valuable
discussion. This research has made use of the NASA/IPAC
Extragalactic Database (NED), which is operated by the Jet
Propulsion Laboratory, California Institute of Technology, under
contract with the National Aeronautics and Space Administration.
We anknowledge the use of public data from SDSS DR9. The SDSS web
site is http://www.sdss.org. Funding for the SDSS has been
provided by the Alfred P. Sloan Foundation, the National
Aeronautics and Space Administration, the National Science
Foundation, the U.S. Department of Energy, and the Japanese
Monbukagakusho, the Max-Planck Society.

\end{acknowledgements}



\label{lastpage}

\end{document}